\documentclass[final,3pt,times]{elsarticle}
\usepackage{amssymb}

\usepackage{amsfonts}

\usepackage{amssymb}
\usepackage{amsmath}
\usepackage{mathrsfs}
\usepackage{graphics}
\usepackage{epsfig}
\usepackage{epstopdf}
\usepackage{amsthm}
\usepackage{textcomp}
\usepackage{color}
\usepackage{float}
\usepackage{multirow}

\usepackage{float}
\usepackage{graphicx}
\usepackage{subfigure}
\usepackage{amsmath,amssymb,amsthm}
\usepackage{amsmath}
\usepackage{amssymb}
\usepackage{epstopdf}
\usepackage{longtable}
\usepackage{rotating}
\usepackage{multirow}
\usepackage{geometry}
\usepackage{caption}
\usepackage{subfigure}

\journal{PLoS One}

\begin{document}

\begin{frontmatter}


\cortext[cor1]{Corresponding author}

\title{The damage inflicted by a computer virus: A new estimation method}


\author[label1]{Jichao Bi}
\ead{272573022@qq.com}

\author[label2,label3]{Lu-Xing Yang}
\ead{ylx910920@gmail.com}

\author[label1]{Xiaofan Yang\corref{cor1}}
\ead{xfyang1964@gmail.com}

\author[label1]{Yingbo Wu}
\ead{wyb@cqu.edu.cn}



\author[label4]{Yuan Yan Tang}
\ead{yytang@umac.mo}

\address[label1]{College of Software Engineering, Chongqing University, Chongqing, 400044, China}

\address[label2]{Faculty of Electrical Engineering, Mathematics and Computer Science, Delft University of Technology, Delft, GA 2600, The Netherlands}

\address[label3]{School of Mathematics and Statistics, Chongqing University, Chongqing, 400044, China}

\address[label4]{Department of Computer and Infomation Science, The University of Macau, Macau}

\begin{abstract}
This paper addressed the issue of estimating the damage caused by a computer virus. First, an individual-level delayed SIR model capturing the spreading process of a digital virus is derived. Second, the damage inflicted by the virus is modeled as the sum of the economic losses and the cost for developing the antivirus. Next, the impact of different factors, including the delay and the network structure, on the damage is explored by means of computer simulations. Thereby some measures of reducing the damage of a virus are recommended. To our knowledge, this is the first time the antivirus-developing cost is taken into account when estimating the damage of a virus.
\end{abstract}

\begin{keyword}
computer virus \sep damage \sep economic losses \sep antivirus cost \sep individual-level delayed SIR model

\end{keyword}

\end{frontmatter}



\section{Introduction}

Computer networks and online social networks provide us the channel of fast information acquisition. Meanwhile, computer viruses can also spread rapidly through these networks, inflicting enormous economic losses \cite{SzorP2005}. As the major means of defending against digital viruses, antivirus programs are capable of detecting and cleaning up viruses within infected hosts, but are incompetent to contain virus spreading in networks. As an emerging interdiscipline, computer virus spreading dynamics aims to explore propagating laws of digital infections by use of the dynamic modeling technique of infectious diseases \cite{MaZE2009}. Since the seminal work by Kephart and White \cite{Kephart1991}, large numbers of computer virus spreading models, ranging from the population-level spreading models \cite{Mishra2011, SongLP2012, YangLX2015d} and the network-level spreading models \cite{Satorras2001, YangLX2014b, RenJG2016a, LiuWP2016a, YangLX2017a} to the individual-level spreading models \cite{Mieghem2009, Sahneh2013, YangLX2015c, YangLX2017b, YangLX2017c}, have been proposed. In particular, the spreading process of a single computer virus is commonly described by Susceptible-Infected-Recovered (SIR) models \cite{Piqueira2009, Rey2013, Ozturk2015}. Additionally, there are some virus spreading models that incorporate the time delay from the appearance of a virus to the release of an antivirus against the virus \cite{YaoY2013a, ZhangZ2015, LiuJ2016}.

One major concern of computer virus spreading dynamics is to estimate the overall damage inflicted by a virus. The overall damage is composed of two parts: the economic losses incurred by the virus and the cost for developing an antivirus against the virus. For the estimation of the economic losses, see Refs. \cite{Khouzani2012, Eshghi2014, YangLX2016, Nowzari2016, BiJC2017}. The cost for developing the antivirus includes the cost for estimating the size of the antivirus to be produced, the cost for estimating the effort required, the cost for developing preliminary project schedules, the cost for estimating overall cost of the project, and the cost for producing the antivirus. The estimation of the antivirus development cost is one of the most challenging tasks in antivirus project management \cite{Boraso1996, Mittal2010}. In real scenarios, the release of the antivirus always lags behind the appearance of the targeted computer virus, and the time delay has a significant influence on the overall damage. When the delay is large, the users of infected hosts would suffer from huge economic losses. However, a reduction of the delay can only be achieved at the expense of a higher antivirus development cost, because it would consume more manpower and financial resources. Therefore, an elaborate tradeoff between the economic losses and the antivirus development cost must be made, so as to minimize the overall damage caused by the digital virus. To our knowledge, previous literatures on computer virus have never taken the antivirus development cost into account.

This paper addressed the issue of estimating the overall damage of a computer virus. First, an individual-level delayed SIR model capturing the spreading process of a digital virus is proposed. Second, the overall damage of the virus is quantified. Next, the impact of different factors, including the delay and the network structure, on the virus damage is uncovered experimentally, thereby some measures of reducing the overall damage are recommended. To our knowledge, this is the first time the antivirus cost is taken into account when estimating the damage of a virus.

The subsequent materials of this work are organized as follows. Section 2 models the damage of a computer virus. Section 3 experimentally explores the influence of different factors on the damage. Finally, Section 4 closes this work.

\newtheorem{rk}{Remark}

\newtheorem{thm}{Theorem}
\newtheorem{lm}{Lemma}
\newtheorem{exm}{Example}
\newtheorem{cor}{Corollary}
\newtheorem{de}{Definition}
\newtheorem{cl}{Claim}
\newtheorem{pb}{Problem}

\newtheorem{con}{Conjecture}
\newtheorem*{pf}{Proof}

\section{Measuring the overall damage}

This section aims to model the damage inflicted by a computer virus. For that purpose, let us introduce some notions, notations and hypotheses as follows.

\subsection{Notions, notations and hypotheses}

Suppose the network in concern consists of $N$ hosts labelled $1, 2, \cdots\, N$. Let $\mathbf{A} = \left(a_{ij}\right)_{N \times N}$ denote the adjacency matrix of the network, i.e., $a_{ij}$ = 1 or 0 according as host $j$ can directly infect host $i$ or not. Suppose a virus appears in the network at time $t = 0$, and there is a delay of $\tau$ time units from the appearance of the virus to the release of an antivirus against the virus. The task of this paper is to estimate the damage inflicted by the virus in the time interval $[0, T]$ , where $T > \tau$.

As with the traditional SIR model, it is assumed that, at any time, every host in the network is in one of three possible states: \emph{susceptible}, \emph{infected}, and \emph{recovered}. Susceptible hosts are hosts that are not infected with the virus but are susceptible to it, because they have not acquired the antivirus. Infected hosts are hosts that are infected with the virus. Recovered hosts are hosts that are not infected with the virus and are immune to it, because they have acquired the antivirus. Let $X_i(t)$ = 0, 1, and 2 denote that, at time $t$, host $i$ is susceptible, infected, and recovered, respectively. Let $S_i(t)$, $I_i(t)$, and $R_i(t)$ denote the probability that host $i$ is susceptible, infected, and recovered at time $t$, respectively. That is,
 \[
 S_i(t) = \Pr\{X_i(t) = 0\}, \quad I_i(t) = \Pr\{X_i(t) = 1\}, \quad R_i(t) = \Pr\{X_i(t) = 2\}.
 \]
As $S_i(t) + I_i(t) + R_i(t) \equiv 1$, the vector
\[
\left(I_1(t), \cdots, I_N(t), R_1(t), \cdots, R_N(t)\right)^T
\]
captures the state of the network at time $t$. Next, let us impose a set of hypotheses as follows.

\begin{enumerate}
    \item[(H$_1$)] Due to the infection by neighboring infected hosts, at time $t \in [0, T]$ a susceptible host $i$ gets infected at rate $\beta\sum_{j=1}^Na_{ij}I_j(t)$, where $\beta > 0$ is referred to as the \emph{infection force}.
    \item[(H$_2$)] Due to the action of the antivirus, at time $t \in [\tau, T]$ every infected host  becomes recovered at rate $\gamma$, where $\gamma > 0$ is referred to as the \emph{recovery rate}.
    \item[(H$_3$)] Due to the action of the antivirus, at time $ t \in [\tau, T]$ every susceptible host becomes recovered at rate $\theta$, where $\theta > 0$ is referred to as the \emph{vaccination rate}.
    \item[(H$_4$)] The loss per unit time suffered by each infected host is one unit.
    \item[(H$_5$)] Due to that the cost for developing an antivirus against the virus goes up sharply when $\tau$ approaches zero, the antivirus development cost is $\frac{A}{\tau^{\alpha}}$, where $A > 0$ is referred to as the \emph{cost coefficient}, $\alpha > 0$ is referred to as the \emph{cost index}.
\end{enumerate}

Fig. 1 shows hypotheses (H$_1$)-(H$_3$) schematically.

\begin{figure}[H]
	\centerline{\includegraphics[height=7.5cm,width=10.5cm]{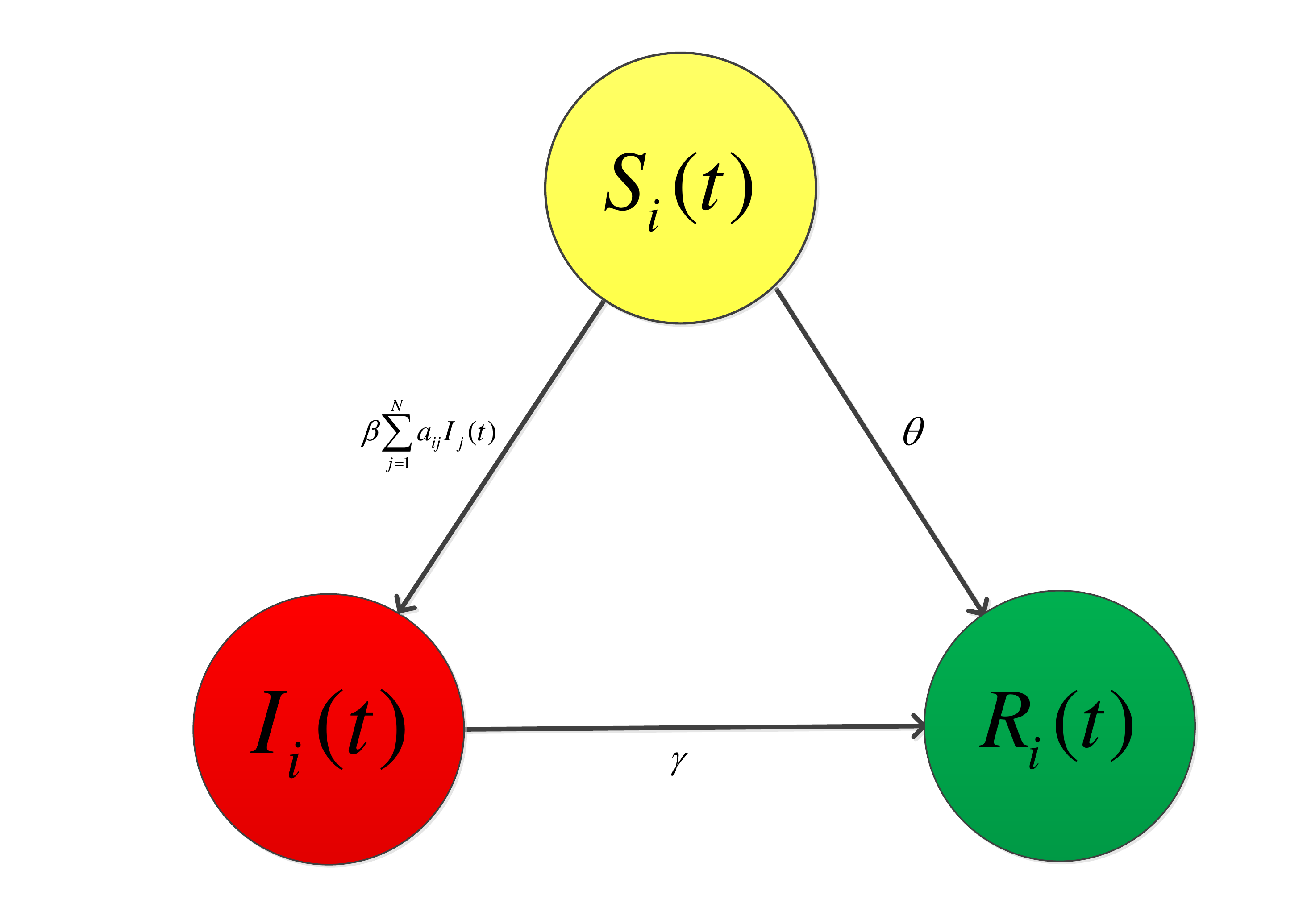}}
	\caption{\textbf{Diagram of hypotheses (H$_1$)-(H$_3$). }}
\end{figure}

\subsection{A delayed SIR model}

Let $\Delta t$ be a very small time interval. For $1 \leq i \leq N$, hypotheses (H$_1$)-(H$_3$) imply the following equations.
\[
  \begin{aligned}
    \Pr\{X_i(t + \Delta t) = 1 \mid X_i(t) = 0\} &= \beta\Delta t \sum_{j=1}^Na_{ij}I_j(t) + o(\Delta t), & 0 \leq t \leq T, \\
    \Pr\{X_i(t + \Delta t) = 2 \mid X_i(t) = 0\}&= 0, &\quad 0 \leq t < \tau - \Delta t,\\
    \Pr\{X_i(t + \Delta t) = 2 \mid X_i(t) = 0\}&= \theta\Delta t + o(\Delta t), &\quad \tau - \Delta t \leq t \leq T,\\
    \Pr\{X_i(t + \Delta t) = 2 \mid X_i(t) = 1\}&= 0, &\quad 0 \leq t < \tau - \Delta t, \\
    \Pr\{X_i(t + \Delta t) = 2 \mid X_i(t) = 1\}&= \gamma\Delta t + o(\Delta t), &\quad \tau - \Delta t \leq t \leq T.
  \end{aligned}
\]
So, for $1 \leq i \leq N$, we have
\[
\begin{aligned}
\Pr\{X_i(t + \Delta t) = 0 \mid X_i(t) = 0\} &= 1 - \beta\Delta t \sum_{j=1}^Na_{ij}I_j(t) + o(\Delta t), & 0 \leq t < \tau - \Delta t, \\
\Pr\{X_i(t + \Delta t) = 0 \mid X_i(t) = 0\} &= 1 - \beta\Delta t \sum_{j=1}^Na_{ij}I_j(t) - \theta\Delta t + o(\Delta t), & \tau - \Delta t \leq t \leq T, \\
\Pr\{X_i(t + \Delta t) = 1 \mid X_i(t) = 1\}&= 1, & 0 \leq t < \tau - \Delta t,\\
\Pr\{X_i(t + \Delta t) = 1 \mid X_i(t) = 1\}&= 1 - \gamma\Delta t + o(\Delta), & \tau - \Delta t \leq t \leq T.
\end{aligned}
\]
By the total probability formula, for $1 \leq i \leq N$, we have
\[
\begin{aligned}
I_i(t + \Delta t)
&= [1 - I_i(t)]\beta\Delta t \sum_{j=1}^Na_{ij}I_j(t) + I_i(t) + o(\Delta t), &\quad 0 \leq t < \tau - \Delta t, \\
I_i(t + \Delta t)
&= [1 - I_i(t) - R_i(t)]\beta\Delta t \sum_{j=1}^Na_{ij}I_j(t) + I_i(t)(1 - \gamma\Delta t) + o(\Delta t), &\quad \tau - \Delta t \leq t \leq T, \\
R_i(t + \Delta t)
&= 0, &\quad 0 \leq t < \tau - \Delta t, \\
R_i(t + \Delta t)
&= [1 - I_i(t) - R_i(t)]\theta\Delta t + I_i(t)\gamma\Delta t + R_i(t) + o(\Delta t), &\quad \tau - \Delta t \leq t \leq T.
\end{aligned}
\]
Transposing, dividing both sides by $\Delta t$, and letting $\Delta t \rightarrow 0$, we get the following dynamical model.
\[
	\left\{
	\begin{aligned}
	  \frac{dI_i(t)}{dt} &= \beta[1 - I_i(t)]\sum_{j=1}^Na_{ij}I_j(t), &\quad 0 \leq t < \tau, 1 \leq i \leq N, \\
      \frac{dI_i(t)}{dt}&= \beta[1 - I_i(t) - R_i(t)]\sum_{j=1}^Na_{ij}I_j(t) - \gamma I_i(t), &\quad \tau \leq t \leq T, 1 \leq i \leq N, \\
      \frac{dR_i(t)}{dt} &= 0, &\quad 0 \leq t < \tau, 1 \leq i \leq N, \\
      \frac{dR_i(t)}{dt}&=\theta[1 - I_i(t) - R_i(t)]+\gamma I_i(t), &\quad \tau \leq t \leq T, 1 \leq i \leq N.
	\end{aligned}
	\right.
\]
We refer to the model as the \emph{delayed SIR model}.

\subsection{Measuring the overall damage}

By hypothesis (H$_4$), the expected economic loss caused by the virus is
\[
\int_0^T\sum_{i=1}^N I_i(t)dt,
\]
On the other hand, it follows from hypothesis (H$_5$) that the antivirus development cost is
\[
\frac{A}{\tau^{\alpha}}.
\]
Hence, the average overall damage of the virus is
\[
  Damage = \int_{0}^T\sum_{i=1}^N I_i(t)dt + \frac{A}{\tau^{\alpha}}.
\]

\section{The influence of different factors on the overall damage}

This section is devoted to exploring the influence of different factors on the overall damage of a computer virus through simulation experiments.

In our experiments, the value ranges of all the model parameters are specified as follows. $\beta \in [0.005, 0.016]$, $\gamma \in [0,1, 0.3]$, $\theta \in [0.1, 0.3]$, $A \in [500, 600]$, $\alpha \in [1, 6]$, and $\tau \in [1, 20]$. The underlying network is taken from a set of five different scale-free networks with 100 nodes, 109 edges, and respective power exponents (2.7, 2.8, 2.9, 3.0, 3.1) \cite{Barabasi1999}, or from a set of five different small-world networks with 100 nodes, 200 edges, and respective edge-rewiring probability (0.1, 0.15, 0.2, 0.25, 0.3) \cite{Watts1998}, or to be a realistic network from the database of Stanford University\cite{stanford}.

\subsection{The influence of the three dynamic parameters}

To understand the influence of the three dynamic parameters (the infection force, the treatment rate, and the vaccination rate) on the overall damage, we present Fig. 2, where each data point is obtained by averaging over $10^4$ runs of the delayed SIR model with different parameter combinations and on different scale-free networks (or different small-world networks, or the realistic network). Thereby, the following conclusions are drawn.

\begin{enumerate}
	\item[(a)] With the rise of the infection force, the overall damage goes up.
	\item[(b)] With the rise of the treatment rate, the overall damage goes down.
	\item[(c)] With the rise of the vaccination rate, the overall damage goes down.
\end{enumerate}

These conclusions manifest that the overall damage caused by a computer virus can be diminished by reducing the infection force, or by enhancing the treatment rate, or by enhancing the vaccination rate. In practical applications, a host user can reduce the infection force of computer viruses by avoiding taking dangerous actions such as browsing suspicious web pages and opening suspicious email attachments, and can enhance the treatment/vaccination rate by timely updating and running antivirus software.

\begin{figure}[H]
   \subfigure{\includegraphics[width=0.31\textwidth,height=4cm]{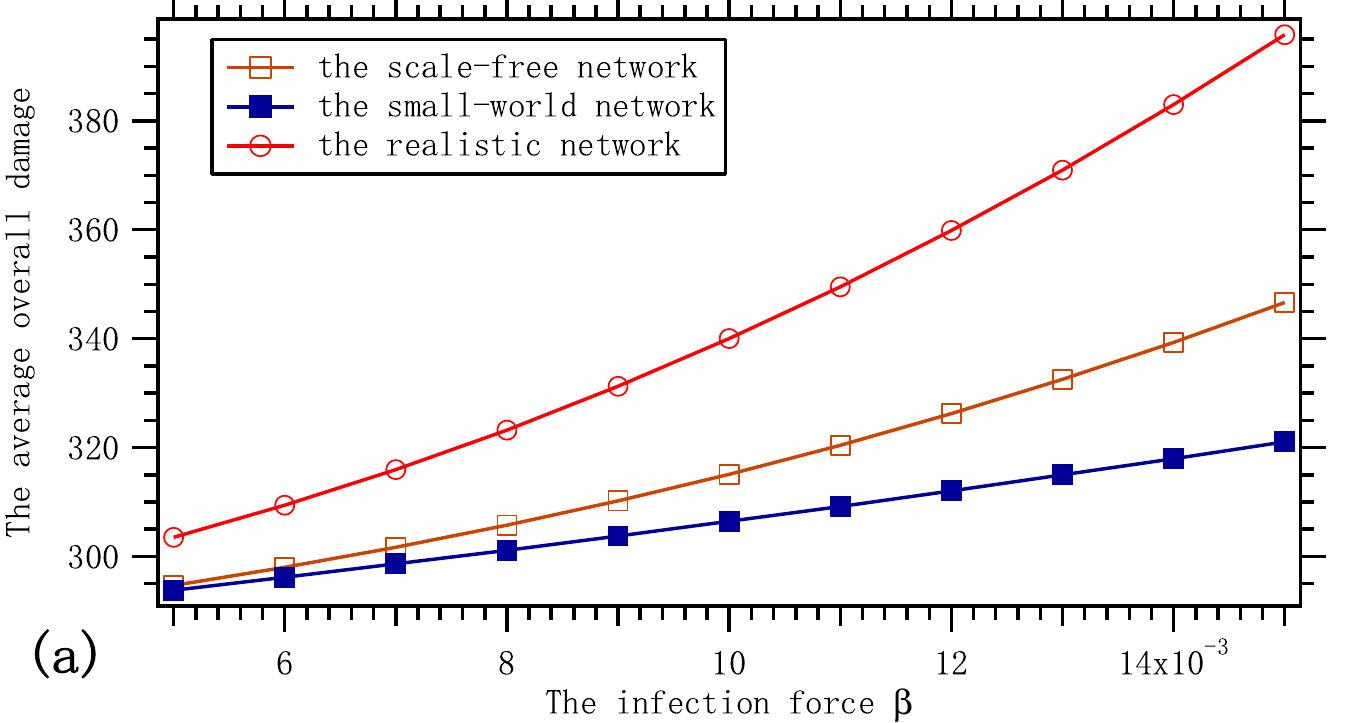}
   \label{fig:a} }
   \subfigure{\includegraphics[width=0.31\textwidth,height=4cm]{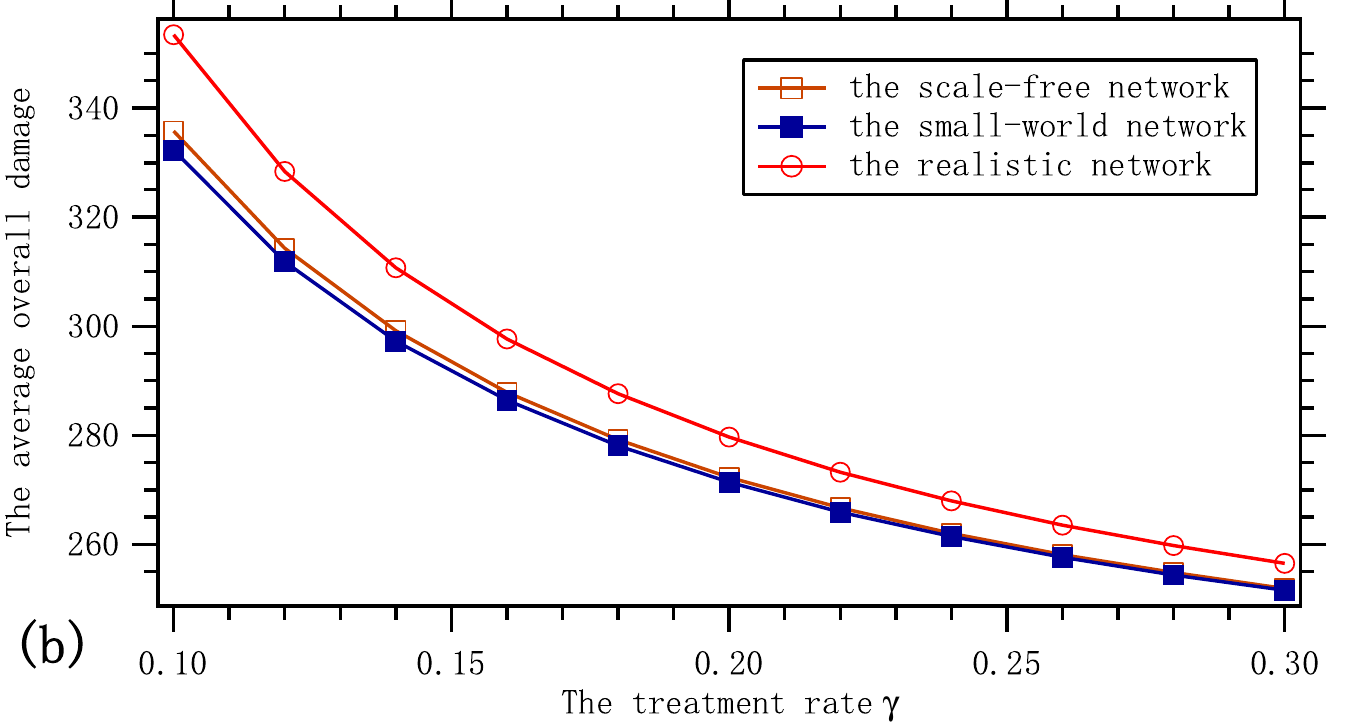}
   \label{fig:b} }
   \subfigure{\includegraphics[width=0.31\textwidth,height=4cm]{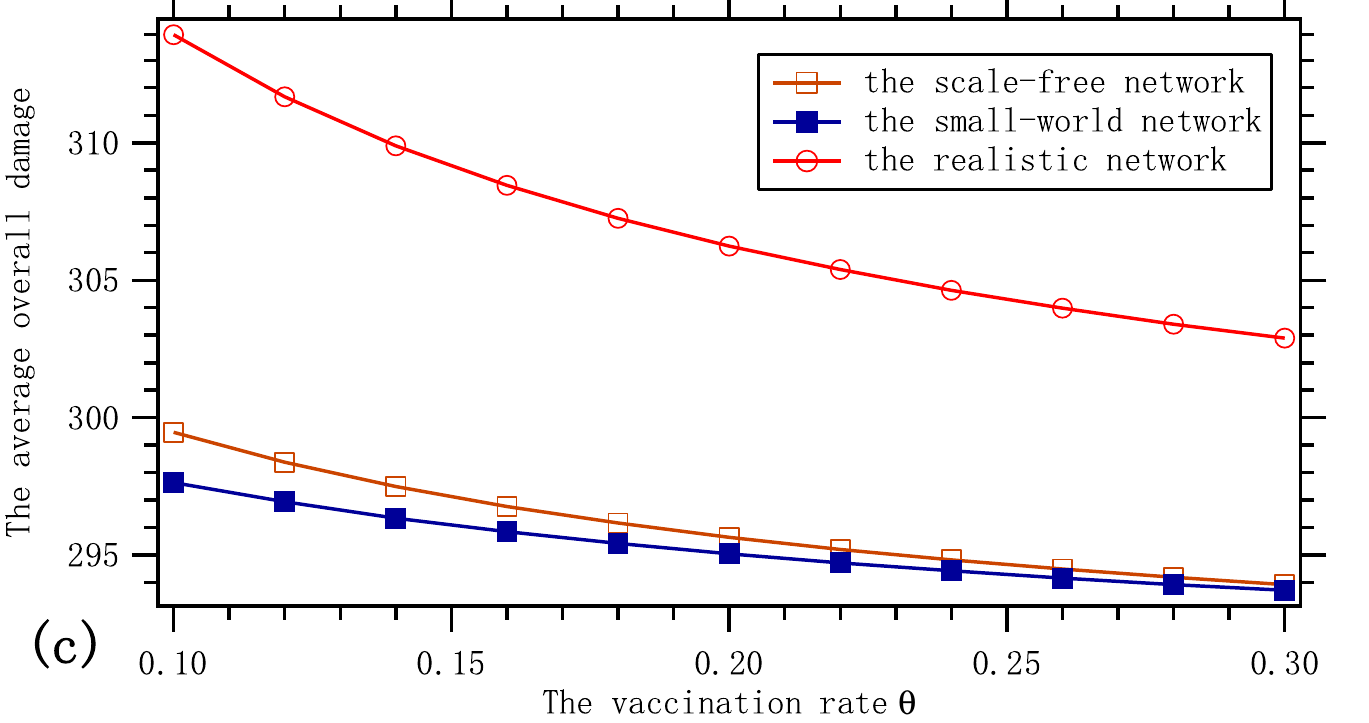}
   \label{fig:c} }
   \caption{\textbf{The average overall damage vs. the three dynamic parameters.} Each data point is obtained by averaging over $10^4$ runs of the delayed SIR model with different parameter combinations and on different scale-free networks (or different small-world networks, or the realistic network).}
\end{figure}

\subsection{The influence of the two cost parameters}

To understand the influence of the two cost parameters (the cost coefficient and the cost exponent) on the overall damage, we present Fig. 3, where each data point is obtained by averaging over $10^4$ runs of the delayed SIR model with different parameter combinations and on different scale-free networks (or different small-world networks, or the realistic network). Thereby, the following conclusions are drawn.

\begin{enumerate}
	\item[(a)] With the rise of the cost coefficient, the overall damage goes up.
	\item[(b)] With the rise of the cost exponent, the overall damage goes down.
\end{enumerate}

These conclusions manifest that the overall damage caused by a computer virus can be diminished by reducing the cost coefficient or by enhancing the cost exponent. In real world applications, the cost for developing an antivirus can be reduced by accurately estimating the workload needed for the development task and building an excellent team for the antivirus development.

\begin{figure}[H]
   \hspace{8ex}
   \subfigure{\includegraphics[width=0.4\textwidth,height=4.5cm]{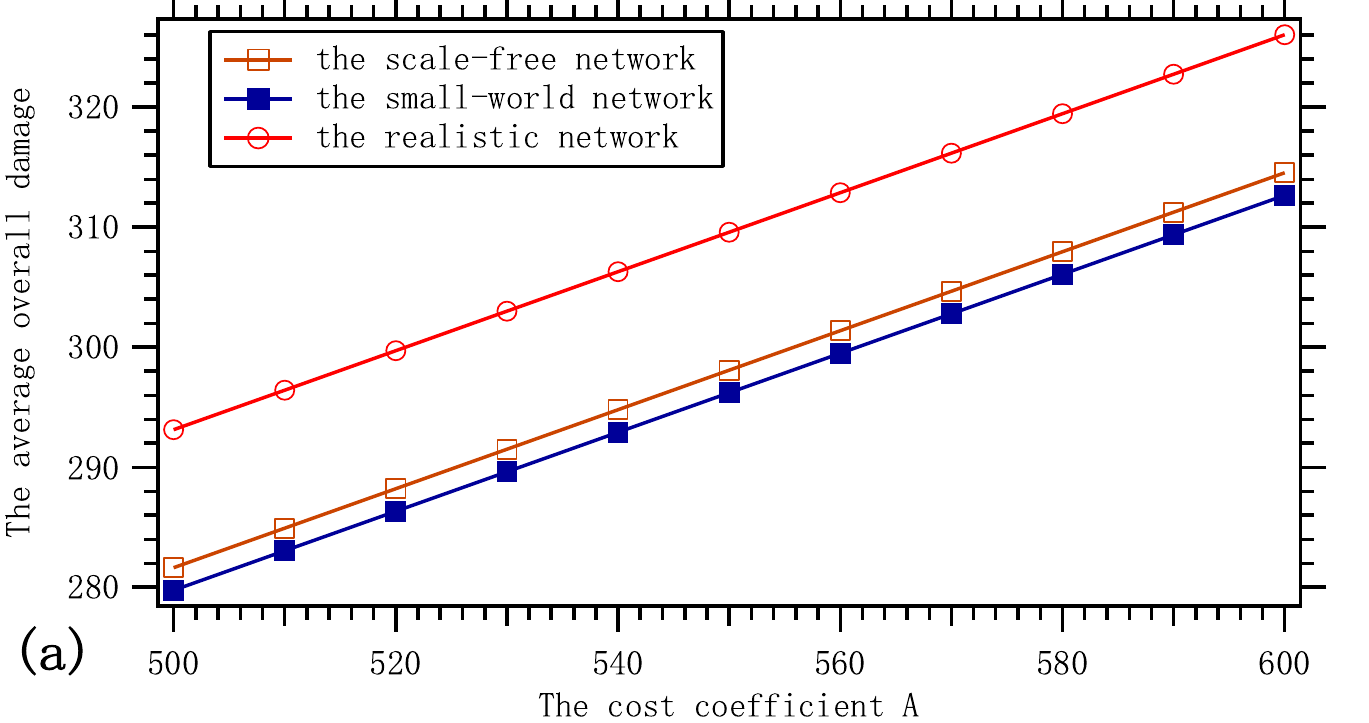}
   \label{fig:a} }
   \hspace{2ex}
   \subfigure{\includegraphics[width=0.4\textwidth,height=4.5cm]{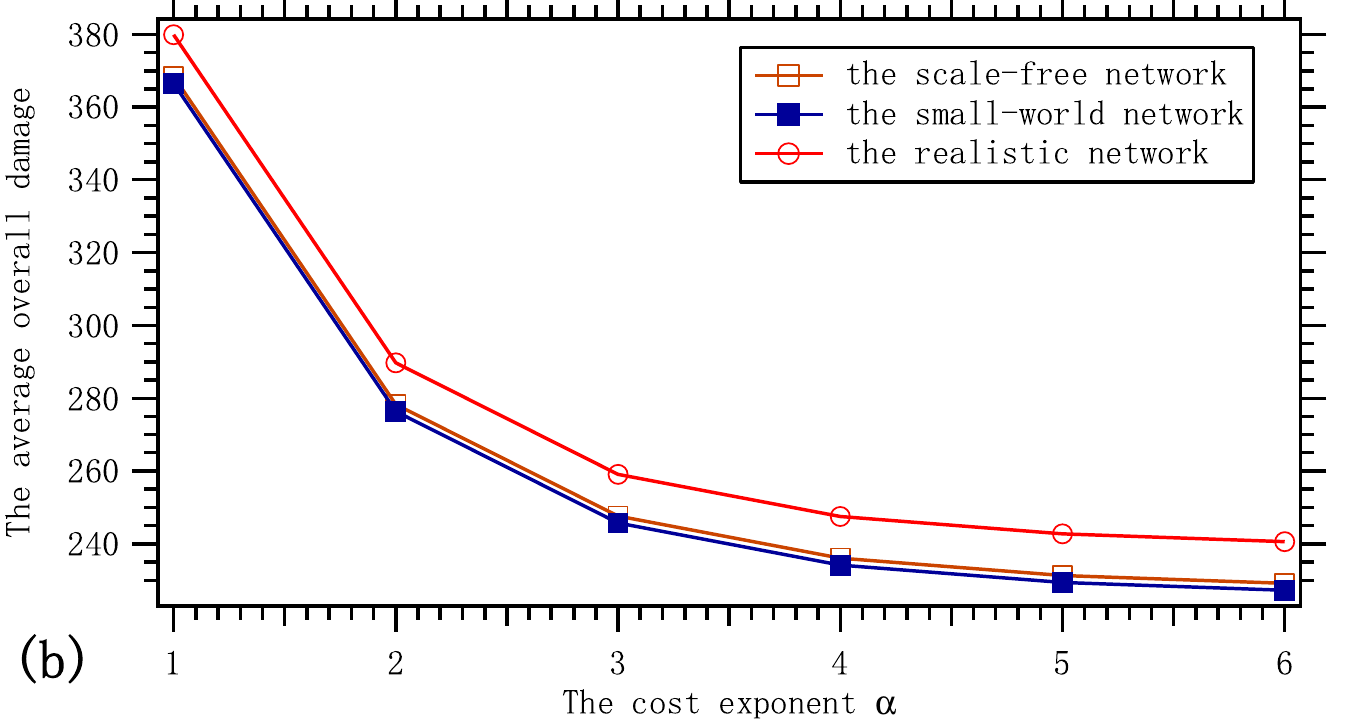}
   \label{fig:a} }
   \caption{\textbf{The average overall damage vs. the two cost parameters.} Each data point is obtained by averaging over $10^4$ runs of the delayed SIR model with different parameter combinations and on different scale-free networks (or different small-world networks, or the realistic network).}
\end{figure}

\subsection{The influence of the time delay}

To understand the influence of the time delay on the overall damage, we present Fig. 4, where each data point is obtained by averaging over $10^4$ runs of the delayed SIR model with different parameter combinations and on different scale-free networks (or different small-world networks, or the realistic network).. Thereby, the following conclusions are drawn.

\begin{enumerate}
\item[(a)] There is a threshold such that (a) when the delay is below the threshold, the overall damage goes down with the rise of the delay, and (b) when the delay exceeds the threshold, the overall damage goes up with the rise of the delay.
\end{enumerate}

In practical uses, an elaborate tradeoff must be made between the economic losses caused by a computer virus and the cost for developing an antivirus against the virus, so as to minimize the overall damage inflicting by the virus.

\begin{figure}[H]
	\centerline{\includegraphics[height=3.5cm,width=6cm]{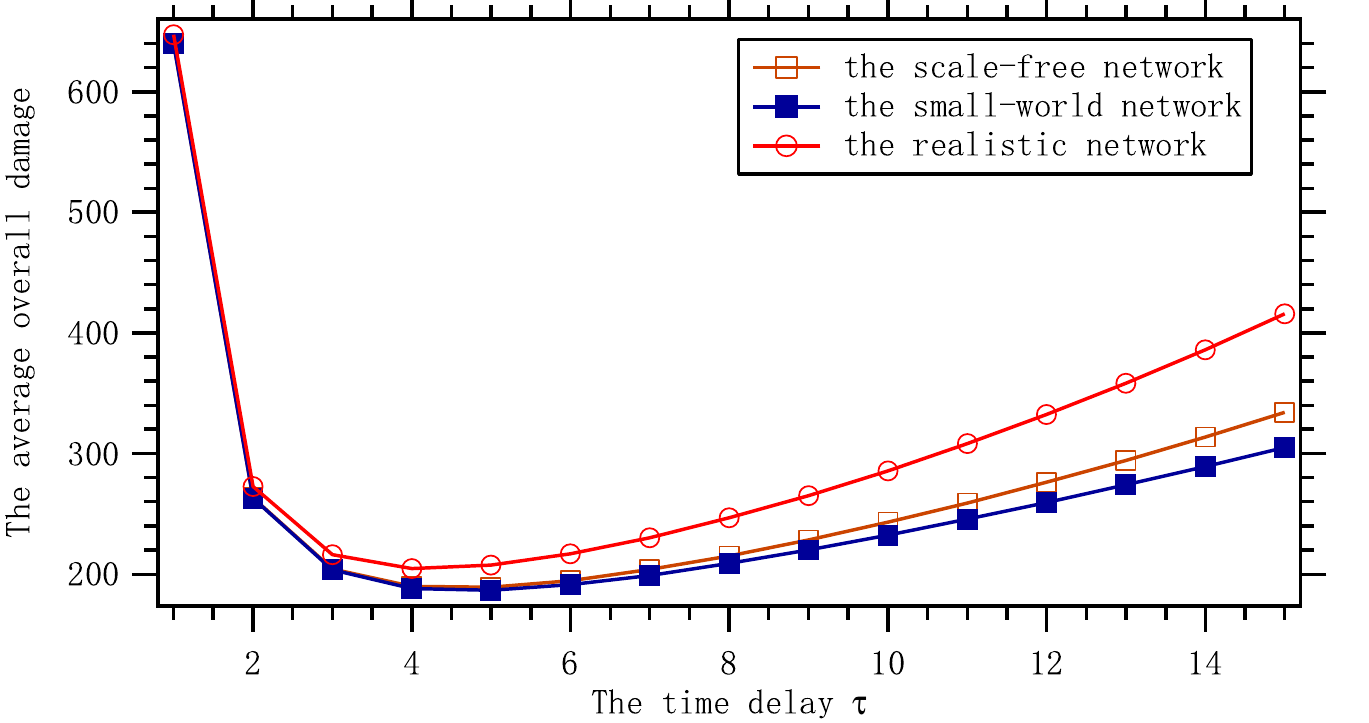}}
	\caption{\textbf{The average overall damage vs. the time delay.} Each data point is obtained by averaging over $10^4$ runs of the delayed SIR model with different parameter combinations and on different scale-free networks (or different small-world networks, or the realistic network).}
\end{figure}

\subsection{The influence of the network structure}

To understand the influence of the network structure on the overall damage, we present Fig. 5, where each data point is obtained by averaging over $10^4$ runs of the delayed SIR model with different parameter combinations and on a specific scale-free network (or a specific small-world network). Thereby, the following conclusions are drawn.

\begin{enumerate}
	\item[(a)] With the rise of the heterogeneity of a scale-free network, the overall damage goes up.
	\item[(b)] With the rise of the randomness of a small-world network, the overall damage goes up.
\end{enumerate}

These conclusions manifest that the overall damage can be reduced by organizing a network in a more homogeneous way or a more regular way. In practice, enterprises can reduce the damage caused by computer viruses by constructing intranets with homogeneous or regular structures.

\begin{figure}[H]
   \hspace{8ex}
   \subfigure{\includegraphics[width=0.4\textwidth,height=4.5cm]{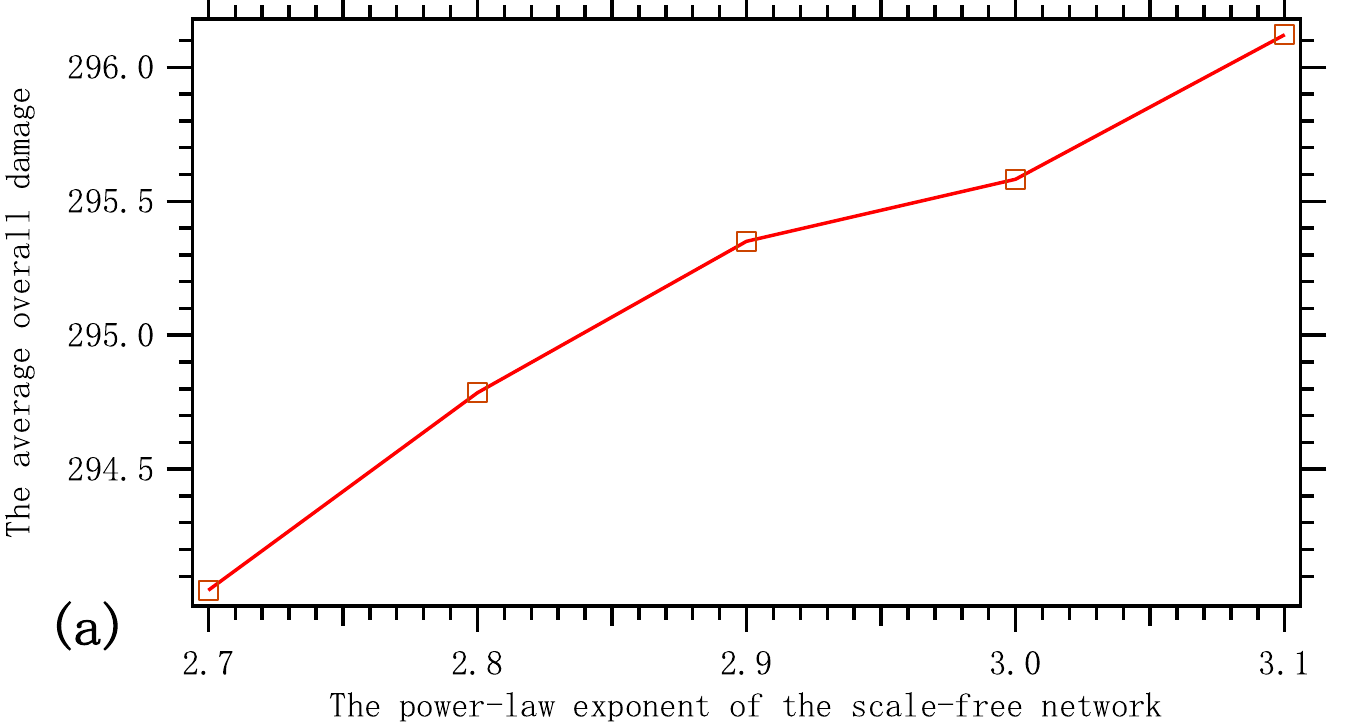}
   \label{fig:a} }
   \subfigure{\includegraphics[width=0.4\textwidth,height=4.5cm]{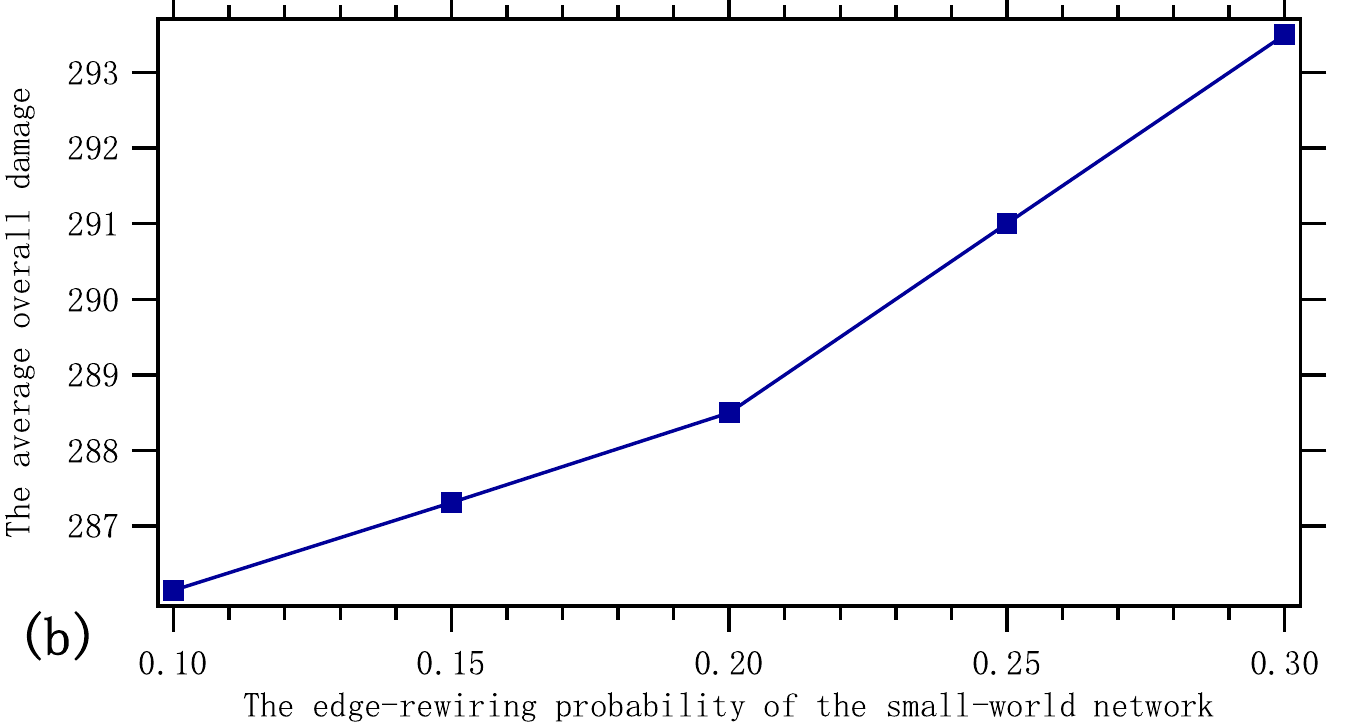}
   \label{fig:b} }
   \caption{\textbf{The average overall damage vs. the network structure.} Each data point is obtained by averaging over $10^4$ runs of the delayed SIR model with different parameter combinations and on a specific scale-free network (or a specific small-world network).}
\end{figure}

\section{Conclusions}

The issue of estimating the overall damage of a computer virus has been addressed. By introducing an individual-level delayed SIR model, the overall damage of the virus has been modeled. The impact of different factors, including the delay and the network structure, on the damage has been uncovered through comprehensive simulation experiments.

Towards this direction, there are still a number of problems that are worth study. For example, the model should be extended to more sophisticated virus spreading models such as the impulsive spreading models \cite{YaoY2012a, YaoY2014, YangLX2014}, the stochastic spreading models \cite{Britton2010, Amador2013, Amador2014}, and the spreading models on time-varying networks \cite{Schwarzkopf2010, Valdano2015, Ogura2015}. As another instance, the methodology developed in this work can be employed to estimate the damage incurred by a rumor \cite{Piqueira2010, Nizamani2013}.


\section*{Acknowledgments}

The authors are grateful to the anonymous reviewers for their valuable suggestions. This work is supported by Natural Science Foundation of China (Grant Nos. 61572006, 71301177), National Sci-Tech Support Plan (Grant No. 2015BAF05B03), Basic and Advanced Research Program of Chongqing (Grant No. cstc2013jcyjA1658), and Fundamental Research Funds for the Central Universities (Grant No. 106112014CDJZR008823).

%
%
%

\section*{References}





\bibliographystyle{elsarticle-num}
\bibliography{<your-bib-database>}



\end{document}